\documentclass[longbibliography, aps, pra, amsmath, amssymb, amsfonts, twocolumn,
superscriptaddress, footinbib]{revtex4-2}

\usepackage{graphicx}
\usepackage{subfigure}
\usepackage{dcolumn}
\usepackage{bm}
\usepackage{bbm}
\usepackage{amsthm}
\usepackage{mathtools}
\usepackage{physics}
\usepackage{binarytree}
\usepackage{tikz}
\usepackage{verbatim}
\usepackage{pgfplots}
\usepackage{placeins}
\usetikzlibrary{decorations.pathreplacing}
\usetikzlibrary{arrows.meta}
\usetikzlibrary{graphs}

\usepackage[colorlinks, linkcolor=red, anchorcolor=blue, citecolor=green]{hyperref}

\pgfplotsset{compat=1.16}
\begin{document}

\title[Quantum teleportation using Ising anyons] {Quantum teleportation using
  Ising anyons}

\author{Cheng-Qian Xu}

\affiliation{Institute of Physics, Beijing National Laboratory for Condensed
  Matter Physics,\\Chinese Academy of Sciences, Beijing 100190, China}

\affiliation{School of Physical Sciences, University of Chinese Academy of
  Sciences, Beijing 100049, China}

\author{D. L. Zhou} \email[]{zhoudl72@iphy.ac.cn}

\affiliation{Institute of Physics, Beijing National Laboratory for Condensed
  Matter Physics,\\Chinese Academy of Sciences, Beijing 100190, China}

\affiliation{School of Physical Sciences, University of Chinese Academy of
  Sciences, Beijing 100049, China}

\affiliation{Collaborative Innovation Center of Quantum Matter, Beijing 100190,
  China}

\affiliation{Songshan Lake Materials Laboratory, Dongguan, Guandong 523808,
  China}

\date{\today}

\begin{abstract}
  Anyons have been extensively investigated as information carriers in
  topological quantum computation. However, how to characterize the information
  flow in quantum networks composed by anyons is less understood, which
  motivates us to study quantum communication protocols in anyonic systems. Here
  we propose a general topologically protected protocol for quantum
  teleportation based on the Ising anyon model, and prove that with our protocol
  an unknown anyonic state of any number of Ising anyons can be teleported from
  Alice to Bob. Our protocol naturally generalizes quantum state teleportation
  from systems of locally distinguishable particles to systems of Ising anyons,
  which may promote our understandings of anyonic quantum entanglement as a
  quantum resource. In addition, our protocol is expected to be realized with
  the Majorana zero modes, one of possible physical realizations for the Ising
  anyon in experiments.
\end{abstract}

                              
\maketitle

\section{Introduction}

Anyon~\cite{PhysRevLett.49.957, PhysRevLett.48.1144}, as a kind of excitations
different from boson and fermion living in two-dimensional system, has attracted
the attention of theorists and experimentalists for its potential applications in
fault-tolerant topological quantum computation due to its non-Abelian braiding
and topologically robustness~\cite{kitaev2003fault, RevModPhys.80.1083, RN8,
  doi:10.1126/science.aaz5601}. In theory, anyon is described by the anyon
model~\cite{KITAEV20062}, which is known as modular tensor category
mathematically~\cite{categoryphysics, 2001Lectures}. One of the most famous is
the Ising anyon model. It is predicted that Majorana zero mode (MZM) as a physical
realization of Ising anyon can exist on some physical platforms, such as
fractional quantum Hall systems~\cite{MOORE1991362, PhysRevLett.111.186401} and
semiconductor nanowires~\cite{Kitaev_2001, doi:10.1126/science.1222360,
  PhysRevLett.105.077001}.

Compared with the quantum information in conventional quantum states of
distinguishable particles, which has a well-established quantum resource
theory~\cite{RevModPhys.91.025001}, the quantum information in anyonic states
is less known to us. Since it is expected that the quantum resource theory of
anyonic states could not only promote the development of topological quantum
computation, but also guide the classification of topological phases in the condensed
matter~\cite{kitaev2006topological, levin2006detecting}. It urges us to study quantum
information theory in anyonic systems. Some effort has being made to investigate
novel quantum resources in anyonic system, such as
anyonic entanglement~\cite{PhysRevA.90.062325, PhysRevB.89.035105, bonderson2017anyonic},
the entropy of anyonic charge entanglement~\cite{bonderson2017anyonic} and
anyonic quantum mutual information~\cite{PhysRevA.104.022610}.

Quantum teleportation~\cite{PhysRevLett.126.090502, naturequantumteleportation}
as the milestone of quantum information and quantum communication, which
utilizes quantum entanglement and classical communication to teleport an unknown
quantum state from one location to another, is a good starting point to investigate
quantum entanglement of anyonic states. As we know the fragility of quantum states
limits the application of quantum teleportation. However,
anyonic system can provide a platform in which quantum states are very robust
to resist environmental interference.
Recently, Huang et al.~\cite{PhysRevLett.126.090502}
have simulated quantum teleportation of a two-MZM state of the Kitaev
chain~\cite{Kitaev_2001} using superconducting qubits, and given a modified
teleportation for teleporting a qubit. It motivates us to consider the
question of whether there is a general protocol for quantum teleportation to teleport
an anyonic state with any number of anyons. However, the answer to this question is
not straightforward. The Hilbert space of anyonic systems doesn't equip with
the tensor product structure like conventional quantum
systems of distinguishable particles~\cite{PhysRevA.69.052326}. Specifically,
the Hilbert space of an anyonic system with total charge $c$ containing
two local subsystems $A$ and $B$ is
\begin{equation}
  \mathcal{H}^c_{AB} = \bigoplus_{a,b} \mathcal{H}_A^a \otimes \mathcal{H}_B^b \otimes V^c_{ab},
\end{equation}
where $a$ ($b$) is the total charge of subsystem $A$ ($B$), and $V^c_{ab}$ is
the fusion Hilbert space associated with the process that two anyons with
charges $a$ and $b$ fuse into an anyon with charge $c$. Thus, traditional
methods such as quantum compiling~\cite{PhysRevLett.125.170501,
  PRXQuantum.2.010334, PhysRevLett.112.140504}, which aims to efficiently
simulate a unitary gate by a series of elementary braidings, cannot be applied
directly.

In this paper, we give a general topological protocol for quantum teleportation using
Ising anyons. Specifically, $2N+1$ copies of the Bell states of two Ising anyons,
\begin{equation}
  \ket{\sigma, \sigma; 0}^{ \otimes (2N+1) },
\end{equation}
which is the maximally entangled state of $4N+2$ Ising anyons~\cite{PhysRevA.104.022610},
are distributed equally to Alice and Bob. Alice can teleport
an anyonic state of $M$ ($M \le 2N+1$) Ising anyons to Bob by their respective
local operation and one-way classical communication. As we all know that the
technical core of the protocol for quantum teleportation is to construct a basis
transformation between the computational basis and the maximally entangled
basis. In the standard quantum teleportation~\cite{nielsen2002quantum,
  RevModPhys.74.347}, Hadamard and CNOT gates are used to play a role
of this basis transformation. In our anyonic teleportation,
however, the key is how to distribute these anyons in an orderly fashion, which
seems trivial to the standard quantum teleportation on the system of distinguishable
particles due to the fact that braiding the distinguishable particles has no effect
on the state. We find a systematic braiding, which equally distributes each
pair of the Bell state of two Ising anyons to Alice and Bob. Based on
this braiding, we give an equation which guides Bob to do the corresponding
local braidings based on Alice's measurement outcomes. This protocol can be
viewed as an extension of quantum teleportation from systems of distinguishable
particles to systems of Ising anyons.

\section{Results}

In this section, we will give a brief review of the Ising anyon
model~\cite{KITAEV20062, pachos_2012}, and present the main result.

The Ising anyon model contains three types of particles, labeled by their
topological charges $\mathcal{I} = \{ 0, \sigma, 1 \}$, where particles $0, \sigma, 1$ are
called vacuum, Ising anyon, and fermion respectively. The fusion rules for the
model are given by
\begin{align}
  \label{eq:FusionR}
  \sigma \times \sigma & = 0 + 1, \\
  1 \times \sigma & = \sigma, \\
  1 \times 1 & = 0.
\end{align}
For example, Eq.~\ref{eq:FusionR} means that when two $\sigma$s are fused there are
two possible fusion results $0$ and $1$.

Utilizing the fusion rules, we can define quantum states of an Ising anyonic system.
For a system with two $\sigma$'s, the Hilbert space is spanned by two orthonormal
anyonic states $\ket{\sigma, \sigma; 0}$ and $\ket{\sigma, \sigma; 1}$, which describe these two
$\sigma$'s coming from vacuum $0$ and a fermion $1$, respectively. When there are
more than two $\sigma$'s, we need to specify the order of fusion. Different orders
will give different bases for the Hilbert space. The transformation between
these bases is achieved through the $F$ matrix (see details in
App.~\ref{app:one}).

In addition to the fusion rules, the Ising anyon model also meets the rules of
braiding. Specifically, exchanging two $\sigma$'s in two-dimension space gives a
unitary transformation named the $R$ matrix (see details in App.~\ref{app:one}).

Unfortunately, based on these two unitary matrices $F$ and $R$, the Ising anyon
model is known to be unable to realize universal topological quantum
computing~\cite{CMP2002Freedman}. Further more, it is shown that not all
Clifford gates can be realized by braiding Ising
anyons~\cite{PhysRevA.79.032311}. However,
multi-qubit Pauli gates, which belong to the Clifford gates, are enough for our teleportation protocol.

\begin{figure}
  \includegraphics{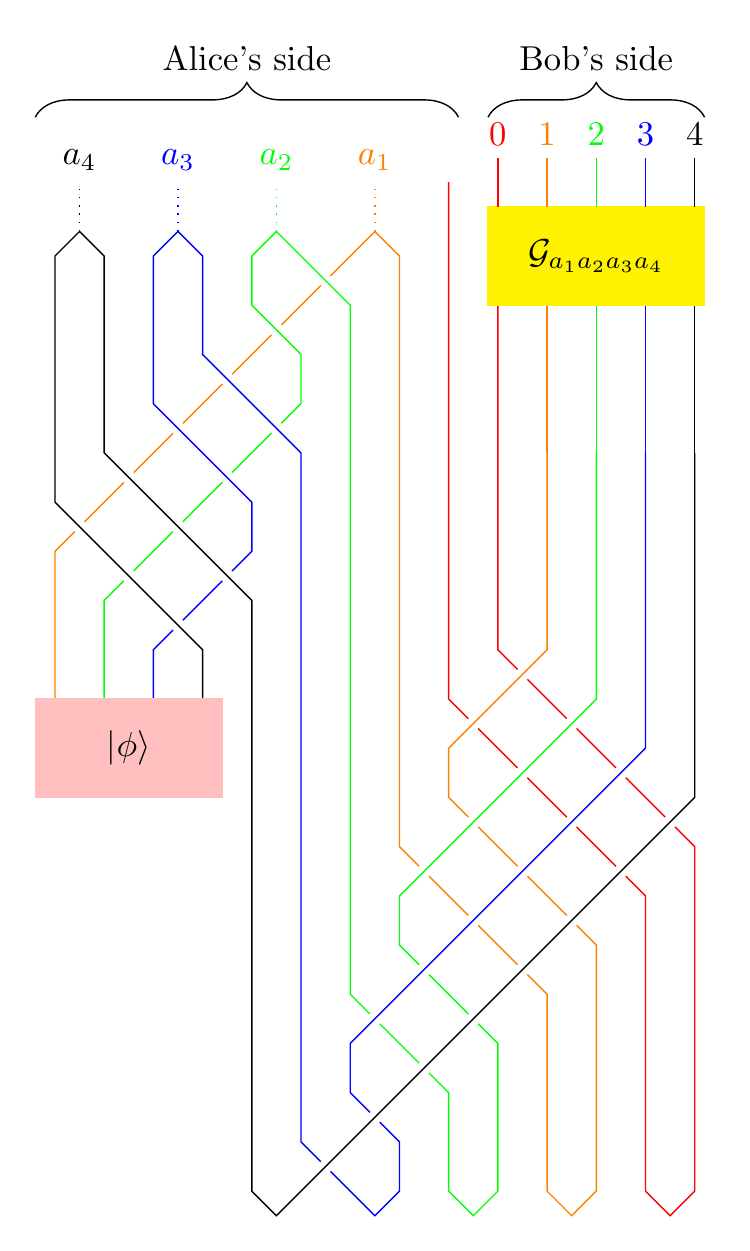}
  \caption{The diagram of the concrete protocol for Ising anyonic teleportation
    via the Bell state of $10$ Ising anyons to teleport an unknown state
    $\ket{\phi}$ of $4$ Ising anyons. First, the Bell state of $10$ Ising anyons
    prepared by the regular braiding is distribute to Alice and Bob. Second,
    Alice performs local braiding to entangle the state $\ket{\phi}$ with the same
    number of anyons of the Bell state. Third, she does projective measurements
    on every neighboring pair, and sends the results
    $\left\{ a_1, a_2, a_3, a_4 \right\}$ to Bob. Fourth, Bob does local
    braidings $\mathcal{G}_{a_1 a_2 a_3 a_4}$ shown in TABLE~\ref{table:1} on
    his $5$ Ising anyons labeled by $0,1,2,3,4$.}
  \label{fig:1}
\end{figure}

Now we are ready to present our protocol for quantum teleportation using Ising
anyons, where a player Alice wants to teleport an unknown state $\ket{\phi}$ of $M$
Ising anyons to a remote player Bob. A non-trivial example with $N = 2$ and
$M = 4$ is illustrated in Fig.~\ref{fig:1}. Our protocol, as a generalization of
the traditional quantum teleportation, is given as four steps:

\textbf{Step 1: Preparation of Ising anyonic Bell state shared by Alice and
  Bob.} The Bell state of $(4N+2)$ Ising anyons [see Eq.~(\ref{eq:bellstate})
for the definition] prepared from $2N+1$ copies of the Bell states of
two Ising anyons $\ket{\sigma, \sigma; 0}^{ \otimes (2N+1) }$,
through braiding the $k$-th copy to the center of the Bell state of $(2k-2)$
Ising anyons in turn [see Eq.~(\ref{eq:Tk})], is shared between Alice and Bob,
where $M \le 2N+1$.

\textbf{Step 2: Ising anyonic Bell state measurement by Alice. } Alice
successively braids the middle two Ising anyons of remaining Ising anyons
to the left side, to entangle the state $\ket{\phi}$ of $M$ Ising anyons with
her $M$ Ising anyons of the Bell state of $(4N+2)$ Ising anyons.
Then, she performs quantum measurements on these $2M$ Ising anyons to obtain
fusion results $a_{k}$ of the $(2k-1)$-th and the $2k$-th Ising anyons.

\textbf{Step 3: Classical communication from Alice to Bob.} Alice sends the
measurement result $\{a_1, a_2,\cdots, a_M\}$ to Bob by one-way classical
communication.

\textbf{Step 4: Conditional operation by Bob.} Bob does braidings
$\mathcal{G}_{a_1 \cdots a_M}$ [see Eqs.~(\ref{eq:geven}) and (\ref{eq:godd})]
on his $M$ Ising anyons based on Alice's measurement outcomes.

We have used the diagrammatic representation for the Ising anyon model in
Fig.~\ref{fig:1}, where solid lines denote Ising anyons, dotted lines denote the
Abelian anyons $\left\{ 0, 1 \right\}$, and the arrow of time goes up from the
bottom. From the figure, we see that the information of the state $\ket{\phi}$,
encoded by $4$ Ising anyonic lines, is accessible by Bob. Here the braiding
performed by Alice and used for the preparation plays a role in building the path
of information flow from the state $\ket{\phi}$ to Bob's output state.

In the following, we will prove the validity of this general protocol. To this
end, we will define the braiding mentioned above named as the tangled braiding,
and give an equation, which guides Bob to perform braidings based on Alice's
measurement outcomes. Finally, the quantum teleportation using Ising anyons will
be built.

\begin{table}
  \begin{tabular}{| c | c | c | c | c |}
    \hline $a_1 a_2 a_3 a_4$ & $\mathcal{G}_{a_1 a_2 a_3 a_4}$ &
    $a_1 a_2 a_3 a_4$ &
    $\mathcal{G}_{a_1 a_2 a_3 a_4}$ \\
    \hline
    $0 0 0 0$ & $I$                        & $1 0 0 0$ & $b_0^2$             \\
    $0 0 0 1$ & $b_0^2 b_1^2 b_2^2 b_3^2$  & $1 0 0 1$ & $b_1^2 b_2^2 b_3^2$ \\
    $0 0 1 0$ & $b_0^2 b_1^2 b_2^2$        & $1 0 1 0$ & $b_1^2 b_2^2$       \\
    $0 0 1 1$ & $b_3^2$                    & $1 0 1 1$ & $b_0^2 b_3^2$       \\
    $0 1 0 0$ & $b_0^2 b_1^2$              & $1 1 0 0$ & $b_1^2$             \\
    $0 1 0 1$ & $b_2^2 b_3^2$              & $1 1 0 1$ & $b_0^2 b_2^2 b_3^2$ \\
    $0 1 1 0$ & $b_2^2$                    & $1 1 1 0$ & $b_0^2 b_2^2$       \\
    $0 1 1 1$ & $b_0^2 b_1^2 b_3^2$        & $1 1 1 1$ & $b_1^2 b_3^2$       \\
    \hline
  \end{tabular}
  \caption{The braidings $\mathcal{G}_{a_1 a_2 a_3 a_4}$ taken by Bob based on
    Alice's measurement outcomes $a_1 a_2 a_3 a_4$. The $b_i$ is the generator
    of the braid, which braids the $i$-th and $(i+1)$-th Ising anyons.}
  \label{table:1}
\end{table}

\section{The tangled braiding}

Before introducing the tangled braiding, let's define the anyonic computational
basis first. Suppose there are $2N$ Ising anyons in a row. We adopt the fusion order
where the $(2k-1)$-th and $2k$-th Ising anyons are fused first. Then, in this
basis the state can be denoted as
\begin{align}
  \ket{a_1 a_2 \cdots a_N; c}_{2N}
  \equiv & \ket{\sigma_1, \sigma_2; a_1} \otimes \ket{\sigma_3, \sigma_4; a_2} \otimes \cdots  \nonumber \\
  & \otimes \ket{\sigma_{2N-1}, \sigma_{2N}; a_N},\label{eq:2}
\end{align}
where $a_i \in \left\{0, 1\right\}$, and $c = \sum_i a_i$ modulo $2$ is the parity.
It should be noted that the parity $c$ is invariant under braiding these $2N$ Ising anyons due to the superselection
rules~\cite{PhysRevA.69.052326}. The computational basis in Eq.~\eqref{eq:2} implies that there is a natural mapping between the Hilbert space of $2N$ Ising anyons and that of $N$ qubits: $\ket{a_1 a_2 \cdots a_N; c}_{2N}$ $\leftrightarrow$ $\ket{a_1 a_2 \cdots a_N}$, where $\ket{a_1 a_2 \cdots a_N}$ is the basis state of $N$ qubits.

Here we are ready to give the definition of the tangled braiding:

\emph{Definition} 1.-- The tangled braiding $\mathcal{T}$ is a series of braidings
acting on $2N$ anyons, which can be written as
$\mathcal{T} = T_N T_{N-1} \cdots T_2$, where braiding $T_k$ ($k \in\{2,3,\cdots,N\}$)
is defined as
\begin{align}
  \label{eq:Tk}
  T_k \equiv
  \begin{tikzpicture}[baseline, scale = 0.6]
    \draw (-1,1) -- (-1,-1);
    \draw (-1.5,1) -- (-1.5,-1);
    \draw (-2.5,1) -- (-2.5,-1);
    \draw[dotted] (1.8,1)--(2.2,1);
    \draw[dotted] (1.3,-1)--(1.7,-1);
    \draw[dotted] (-1.8,1)--(-2.2,1);
    \draw[dotted] (-1.8,-1)--(-2.2,-1);
    \draw (2,-1)--(2.5,1);
    \draw (1,-1)--(1.5,1);
    \draw (0.5,-1)--(1,1);
    \draw (3.5,-1)--(2.3,-0.2); \draw (2.1,-0.07) -- (1.5,0.33);
    \draw (1.25,0.5) -- (1,0.67); \draw (0.9,0.73) -- (0.5,1);
    \draw (2.5,-1) -- (2.15,-0.77); \draw (2,-0.67) -- (1.3,-0.2);
    \draw (1.1,-0.07) -- (0.9,0.07); \draw (0.7,0.2) -- (-0.5,1);
    \draw [decorate,decoration={brace,amplitude=10pt},xshift=0pt,yshift=6pt]
    (1,1)-- node[midway,yshift=0.6cm]{$(k-1)$ $\sigma$'s}(2.5,1);
    \draw [decorate,decoration={brace,amplitude=10pt},xshift=0pt,yshift=-6pt]
    (-1,-1)-- node[midway,yshift=-0.6cm]{$(k-1)$ $\sigma$'s}(-2.5,-1);
  \end{tikzpicture}.
\end{align}

This tangled braiding $\mathcal{T}$ can be viewed as moving the $(2k-1)$-th
and $2k$-th Ising anyons from the right to the $k$-th and $(k+1)$-th positions
respectively, in turn from $k=2$ to $k=N$.

The Bell state of $10$ Ising anyons mentioned in the previous sections is just
the state denoted as $\ket{\mathcal{B}_{10}(00000); 0}$ that can be obtained
through the tangled braiding $\mathcal{T}$,
\begin{align}
  \label{eq:bellstate}
  \ket{\mathcal{B}_{10}(00000); 0} & \equiv \mathcal{T} \ket{00000; 0}_{10} \nonumber \\
  & =\left( \frac{1}{d_\sigma} \right)^{\frac{5}{2}}
  \begin{tikzpicture}[baseline, scale = 0.5]
    \draw (-0.5,1)--(0,0.5)--(0.5,1);
    \draw (-1,1)--(0,0)--(1,1);
    \draw (-1.5,1)--(0,-0.5)--(1.5,1);
    \draw (-2,1)--(0,-1)--(2,1);
    \draw (-2.5,1)--(0,-1.5)--(2.5,1);
  \end{tikzpicture}.
\end{align}
This is the maximally entangled state that maximizes the anyonic von Neumann entropy
$\tilde{S}(\tilde{\rho}) = \tilde{\rm Tr} \left[ \tilde{\rho} {\rm log} \tilde{\rho} \right]$
of the anyonic reduced state $\tilde{\rho}$ of the left (or right) $5$ Ising
anyons, where $\tilde{\rm Tr}$ is the quantum trace~\cite{bonderson2017anyonic}.
In addition to this state, another state that we will use in the following is
\begin{align}
  \ket{\mathcal{B}_{10}(10000); 1} & \equiv \mathcal{T} \ket{10000; 1}_{10} \nonumber \\
  & =\left( \frac{1}{d_\sigma} \right)^{\frac{5}{2}}
  \begin{tikzpicture}[baseline, scale = 0.5]
    \draw (-0.5,1)--(0,0.5)--(0.5,1);
    \draw (-1,1)--(0,0)--(1,1);
    \draw (-1.5,1)--(0,-0.5)--(1.5,1);
    \draw (-2,1)--(0,-1)--(2,1);
    \draw (-2.5,1)--(0,-1.5)--(2.5,1);
    \draw[dotted] (0,-1.5) -- (0,-2.5) node[pos=0.5, right]{$1$};
  \end{tikzpicture}.
\end{align}

To verify the validity of our quantum teleportation protocol,  we prove a useful equality for the tangled braiding.

\emph{Lemma}.-- For the state $\ket{a_1 a_2; a_1+a_2}_{4}$ of $4$ Ising
anyons, applying the tangled braiding $\mathcal{T}$ and then braiding the first two
anyons twice successively, equals applying the braiding $\mathcal{T}$ acting on
the state $\ket{(a_1 + 1) (a_2 + 1); a_1+a_2}_{4}$ up to a global phase:
\begin{align}
  \label{eq:11}
  \begin{tikzpicture}[baseline]
    \draw (0,0) -- (0,-0.5) -- (0.25,-0.75) -- (1.5,0.5) -- (1.5,1.5);
    \draw (0.8,-0.3) -- (1.25,-0.75) -- (1.5,-0.5) -- (1.5,0) -- (1.3,0.2);
    \draw (0.7,-0.2) -- (0.5,0) -- (0.5,0.5) -- (0.3,0.7);
    \draw (0,0) -- (0,0.5) -- (0.5,1) -- (0.3,1.2);
    \draw (0.2,0.8) -- (0,1) -- (0.5,1.5);
    \draw (0.2,1.3) -- (0,1.5);
    \draw (1.2,0.3) -- (1,0.5) -- (1,1.5);
    \draw[dotted] (0.25,-0.75) -- (0.25,-1.25) node[pos=1,below]{$a_1$};
    \draw[dotted] (1.25,-0.75) -- (1.25,-1.25) node[pos=1, below]{$a_2$};
  \end{tikzpicture} \quad =
  \begin{tikzpicture}[baseline]
    \draw (0,0) -- (0,-0.5) -- (0.25,-0.75) -- (1.5,0.5) -- (1.5,1);
    \draw (0.8,-0.3) -- (1.25,-0.75) -- (1.5,-0.5) -- (1.5,0) -- (1.3,0.2);
    \draw (0.7,-0.2) -- (0.5,0) -- (0.5,0.5) -- (0.5,1);
    \draw (0,0) -- (0,1);
    \draw (1.2,0.3) -- (1,0.5) -- (1,1);
    \draw[dotted] (0.25,-0.75) -- (0.25,-1.25) node[pos=1, below]{$a_1 + 1$};
    \draw[dotted] (1.25,-0.75) -- (1.25,-1.25) node[pos=1, below]{$a_2 + 1$};
  \end{tikzpicture}.
\end{align}

\emph{Proof}.-- We check this equality by mapping these braidings to
the corresponding unitary operators. By using the $R$ and $F$ matrices of the Ising
anyon model, we find that all $n$-qubit Pauli gates can be obtained by braiding
$2n+1$ Ising anyons in the anyonic computational basis for $2n+2$ Ising anyons
(see details in App.~\ref{app:two}):
\begin{align}
  & \left( b^{(2n+2)}_{2j-1} \right)^2 = \tau_3^{(j)}, \nonumber \\
  & \left( b^{(2n+2)}_{2j} \right)^2 = \tau^{(j)}_1 \otimes \tau_1^{(j+1)} \nonumber \\
  & b^{(2n+2)}_{2n-1} \left( b^{(2n+2)}_{2n} \right)^2 b^{(2n+2)}_{2n-1} \left( b^{(2n+2)}_{2n} \right)^2 = i, \nonumber
\end{align}
where $1 \le j \le n$, $\tau_i^{(j)}$ is Pauli matrix $\tau_i$ acting on the
$j$-th qubit $a_j$ in the anyonic computational basis, and $b^{(2n+2)}_{j}$ is
the generator of braid group $B_{2n+2}$. Thus, braiding the first two Ising
anyons gives the Pauli gate $\tau_3$ acting on the first qubit $a_1$. And the
tangled braiding $\mathcal{T}$ gives two-qubit entangled gate:
\begin{align}
  \frac{1}{\sqrt{2}}
  \begin{pmatrix}
    1 & 0 & 0 & -i \\
    0 & i & 1 & 0 \\
    0 & -i & 1 & 0 \\
    1 & 0 & 0 & i \\
  \end{pmatrix}.
\end{align}
Using these two gates, we can check Eq.~(\ref{eq:11}) directly.
$\blacksquare$

The physical significance of the tangled braiding $\mathcal{T}$ is apparent,
which distributes the entanglement in $N$ copies of the Bell states of two Ising anyons from left to right between Alice and Bob. Symmetrically, we can also define the dual tangled braiding
$\tilde{\mathcal{T}}$, which distributes the entanglement in $N$ copies from right to left.
 This dual tangled braiding $\tilde{\mathcal{T}}$ is just the tangled braiding $\mathcal{T}$ seen from the other side.
Therefore, we will not distinguish between these two braidings, and use the symbol
$\mathcal{T}$ uniformly.

\section{Ising anyonic teleportation}

Now we will show that, by utilizing the tangled braiding $\mathcal{T}$ given
above, we can teleport an anyonic state of Ising anyons from Alice to Bob, which
can be seen as an anyonic version of quantum teleportation referred as Ising
anyonic teleportation.

\begin{figure}
  \centering \subfigure[]{\label{fig:suba}}
  \includegraphics[width=0.45\linewidth]{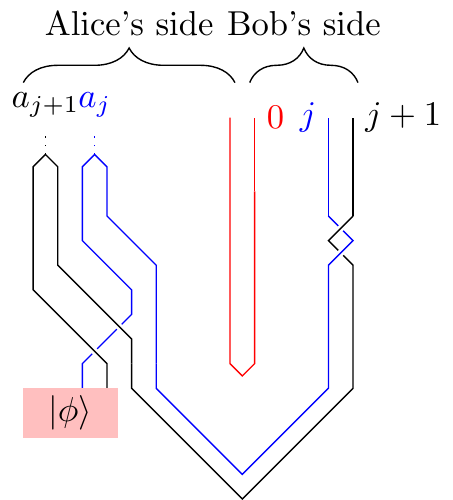} \hspace{0.005\linewidth}
  \subfigure[]{\label{fig:subb}}
  \includegraphics[width=0.45\linewidth]{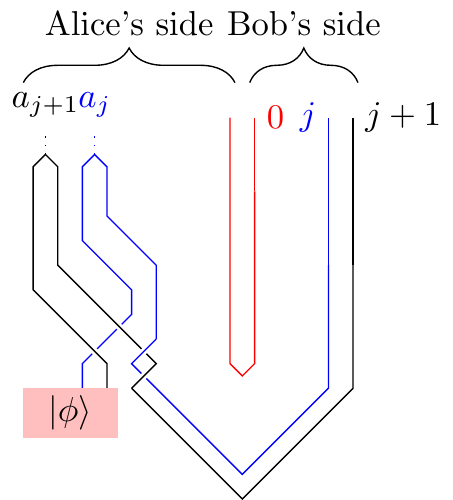}
  \caption{The diagrams for the local braidings applied by Bob can be
    transmitted to Alice's side. The endpoint of blue (black) line on the right
    side denotes the $j$-th ($(j+1)$-th) Ising anyon owned by Bob, which is
    connected to the measurement outcome $a_j$ ($a_{j+1}$) on Alice's side. The
    red line is the Bell state of two Ising anyons shared between Alice and Bob,
    which is prepared to change the parity $c$ of Alice's measurement outcomes.}
  \label{fig:2}
\end{figure}

\emph{Theorem}.-- In the Ising anyonic teleportation, Alice and Bob share
the state $\ket{\mathcal{B}_{4N+2}(0 \cdots 0); 0}$, which can be prepared from
the state $\ket{0 \cdots 0; 0}_{4N+2}$ by the tangled braiding $\mathcal{T}$,
and each takes $2N+1$ Ising anyons. Alice has another unknown state $\ket{\phi}$ of $M$
($M \le 2N+1$) Ising anyons, which is prepared to be teleported to
Bob. On Alice's side, Alice does the operation $\mathcal{T}^{-1}$ by braiding
her $2M$ Ising anyons, then she measures every neighboring pair of these $2M$
Ising anyons to obtain $M$ fusion results $\left\{ a_1 \cdots a_M \right\}$. She
sends these results to Bob through one-way communication, who performs
corresponding local operation $\mathcal{G}_{a_1 \cdots a_M}$ that depends on
the results of Alice's measurement by braiding his Ising anyons.
When $M$ is even,
\begin{align}
  \label{eq:geven}
  \mathcal{G}_{a_1 \cdots a_M} = & (b_0)^{2c} (b_1)^{2(c - a_1)} \cdots
  (b_j)^{2\sum_{i=j+1}^M a_i} \nonumber \\
  & \cdots ( b_{M-2} )^{2 (a_M + a_{M-1})} ( b_{M-1} )^{2 a_M}.
\end{align}
When $M$ is odd,
\begin{align}
  \label{eq:godd}
  \mathcal{G}_{a_1 \cdots a_M} = & (b_1)^{2(c - a_1)} \cdots (b_j)^{2\sum_{i=j+1}^M a_i} \nonumber \\
  & \cdots ( b_{M-2} )^{2 (a_M + a_{M-1})} ( b_{M-1} )^{2 a_M}.
\end{align}
Here $c = \sum_{i=1}^M a_i$ modulo $2$ is the parity, $b_j$ is the generator of
Braid group, which braids the $j$-th and $(j+1)$-th anyons
(see Fig.~\ref{fig:2} with $M=4$).
Finally, the original anyonic state $\ket{\phi}$
owned by Alice will appear on Bob's side.

Specifically, in the Theorem, the inverse of the tangled braiding $\mathcal{T}^{-1}$
performed by Alice on $2M$ Ising anyons can be broken down into a series of
fairly simple braidings:
\begin{equation}
  \mathcal{T}^{-1} = T_2^{-1} T_3^{-1} \cdots T_M^{-1}.
\end{equation}
From Eq.~(\ref{eq:Tk}), we know that $T_k$ is the operation that moves
the $(2k-1)$-th and $2k$-th Ising anyons to the $k$-th and $(k+1)$-th positions
respectively. The inverse of $T_k$ is the opposite operation that moves
the $(k+1)$-th and $k$-th Ising anyons to the $2k$-th and $(2k-1)$-th positions
respectively. An example with $M=4$ is illustrated in Fig.~\ref{fig:1}.
We can see that Alice first performs $T_4^{-1}$ (black lines) that moves
the $5$-th and $6$-th Ising anyons to the $8$-th and $7$-th positions respectively,
and then performs $T_3^{-1}$ (blue lines) and $T_2^{-1}$ (green lines).
The operation $\mathcal{T}^{-1}$ pairs the Ising anyons of the state $\ket{\phi}$
with the Ising anyons of the shared Bell state in an orderly fashion.

\emph{Proof}.--The main idea of the proof is to show that, for Bob,
the case that Alice's measurement outcomes are not $\{ 0\cdots 0 \}$ can
be equivalent to the case with outcomes $\{ 0\cdots 0 \}$ by his operation.

First we note that, when Alice's measurement outcomes are
$\left\{ 0 \cdots 0 \right\}$, Bob will obtain the $\ket{\phi}$ by doing
nothing. Since we have
\begin{align}
  \label{eq:zigzag}
  \begin{tikzpicture}[baseline, scale = 0.5]
    \draw (0.5,2) -- (0.5,0) -- (0,-0.5) -- (-2.5,2) -- (-4.5,0) -- (-4.5,-0.5);
    \draw (1,2) -- (1,0) -- (0,-1) -- (-2.5,1.5) -- (-4,0) -- (-4,-0.5);
    \draw (2,2) -- (2,0) -- (0,-2) -- (-2.5,0.5) -- (-3,0) -- (-3,-0.5);
    \draw (-4.7,-0.5) rectangle (-2.8,-1.5);
    \draw (-3.75,-1) node[]{$\ket{\phi}$};
    \draw[dotted] (-3.7,0) -- (-3.3,0);
    \draw[dotted] (-2.5,1.2) -- (-2.5,0.8);
    \draw[dotted] (0,-1.3) -- (0,-1.7);
    \draw[dotted] (1.3,0) -- (1.7,0);
  \end{tikzpicture} \; =
  \begin{tikzpicture}[baseline, scale = 0.6]
    \draw (3.8,-0.5) rectangle (5.7,-1.5);
    \draw (4,-0.5) -- (4,2);
    \draw (4.5,-0.5) -- (4.5,2);
    \draw (5.5,-0.5) -- (5.5,2);
    \draw[dotted] (4.8,0) -- (5.2,0);
    \draw (4.75,-1) node[]{$\ket{\phi}$};
  \end{tikzpicture},
\end{align}
which follows the fact that the zigzag gives
identity~\cite{categoryphysics}. In other words, these two diagrams are
topological equivalent (isotopy) by continuous deformations as long as open
endpoints are fixed~\cite{KITAEV20062}.

Second, for other measurement outcomes $\{ a_1 \cdots a_M \}$, the diagram will
be complicated and can not become the diagram on left-hand side of
Eq.~(\ref{eq:zigzag}) directly.
Bob's target is to deform the diagram to become the diagram on left-hand side
of Eq.~(\ref{eq:zigzag}) by his local operation. Indeed, Bob can realize it
by taking advantage of the Bell state $\ket{\mathcal{B}_{4N+2}(0 \cdots 0); 0}$
shared between Alice and him. To see this, without loss of generality, we consider two
adjacent anyons on Bob's side, namely, the $j$-th and $(j+1)$-th anyons, which
are connected to the measurement outcomes $a_j$ and $a_{j+1}$ on Alice's side as
shown in Fig.~\ref{fig:2}. It is reasonable to discuss only the $j$-th and $(j+1)$-th
anyons here and ignore the other, since braiding on any adjacent anyons will not affect
other anyons which can be seen from Fig.~\ref{fig:1}. We suppose that Bob braids
these two anyons twice illustrated in Fig.~\ref{fig:suba}. Topologically,
we can move this braiding performed by Bob to Alice' side through the Bell state
of $4$ Ising anyons shared between them, whereby we obtain the diagram shown
in Fig.~\ref{fig:subb}. We notice that the diagram on Alice's side
in Fig.~\ref{fig:subb} is exactly the left-hand side of Eq.~(\ref{eq:11}).
Taking advantage of the Lemma in the previous section,
the case that Bob braids the $j$-th and $(j+1)$-th Ising anyons twice for
Alice's measurement outcomes $\{ a_{j}, a_{j+1} \}$ is equivalent to the case
that Bob does nothing for Alice's measurement outcomes $\{ a_{j}+1, a_{j+1}+1 \}$.
We conclude that when Alice's measurement outcome $a_{j+1} = 1$, Bob only
needs to braid the $(j+1)$-th and $j$-th Ising anyons twice on his side, which
is equivalent to the case with $a_{j+1} = 0$.

Take $M=4$ as an example illustrated in TABLE~\ref{table:1}.
When Alice's measurement outcomes are $\{0011\}$, Bob only needs to braid
the $3$-rd  and $4$-th Ising anyons twice, which is equivalent to the case
with measurement outcomes are $\{0000\}$. However, it should be noted that
when the parity $c$ of measurement outcomes is odd, Bob needs to use auxiliary
Ising anyon labeled as $0$-th illustrated in Fig.~\ref{fig:2}.
Thus, Bob can take the local braiding $\mathcal{G}_{a_1 \cdots a_M}$ presented
in Eq.~(\ref{eq:geven}) to hit the mark.

Third, we find that there is no need to perform the last step, $b_{0}^2$, to
change $a_1$ if $M$ is odd. To see it, we give the anyonic state of $M$ (odd)
Ising anyons when Alice's measurement outcomes are $\{ 1 0 \cdots 0 \}$:
\begin{align}
  \label{eq:16}
  \begin{tikzpicture}[baseline, scale = 0.5]
    \draw (0.5,2) -- (0.5,0) -- (0,-0.5) -- (-2.5,2) -- (-4.5,0) -- (-4.5,-0.5)
    -- (-4.5,-1.5) -- (-4,-2) -- (-4,-2.5);
    \draw[dotted] (-3.5,-1.5) -- (-4,-2) node[pos=0.2, below]{$i$};
    \draw[dotted] (-2.5,2) -- (-2.5,3) node[pos=0.5, right]{$c=1$};
    \draw (1,2) -- (1,0) -- (0,-1) -- (-2.5,1.5) -- (-4,0) -- (-4,-0.5);
    \draw (2,2) -- (2,0) -- (0,-2) -- (-2.5,0.5) -- (-3,0) -- (-3,-0.5);
    \draw (-4.2,-0.5) rectangle (-2.8,-1.5);
    \draw (-3.5,-1) node[]{$\ket{\psi}$};
    \draw[dotted] (-3.7,0) -- (-3.3,0);
    \draw[dotted] (-2.5,1.2) -- (-2.5,0.8);
    \draw[dotted] (0,-1.3) -- (0,-1.7);
    \draw[dotted] (1.3,0) -- (1.7,0);
  \end{tikzpicture} \quad = \quad
  \begin{tikzpicture}[baseline, scale = 0.7]
    \draw (-0.2,0) rectangle (1.2,1);
    \draw (0.5,0.5) node[]{$\ket{\psi}$};
    \draw (-0.3,1.2) -- (-1,0.5) -- (0,-0.5) -- (0,-1.5);
    \draw[dotted] (0.5,0) -- (0,-0.5) node[pos=0.2, below]{$i$};
    \draw[dotted] (-1.7,1.2) -- (-1,0.5) node[pos=0, above]{$c$};
    \draw (0,1) -- (0,1.2);
    \draw (1,1) -- (1,1.2);
    \draw[dotted] (0.3,1.2) -- (0.7,1.2);
  \end{tikzpicture} ~,
\end{align}
where $\ket{\psi}$ is the state of $M-1$ (even) Ising anyons with the parity $i$.
On the right-hand side of Eq.(\ref{eq:16}), charge $c$ denotes the total charged obtained by Alice's measurement.
By quantum tracing the charge $c$ (Alice's side), we can obtain the
state $\ket{\phi}$ of $M$ Ising anyons (see App.~\ref{app:three}),
which Alice wants to teleport.

In conclusion, we have shown that this Ising anyonic teleportation works.
$\blacksquare$

In the above proof, we have used the Bell state of $4N+2$ Ising anyons to move
the braiding performed by Bob to Alice side. In quantum information, we have a
similar situation where we can move a unitary gate from one party to the other
by using the general Bell state
$(1/\sqrt{d})\sum_i \ket{i}_A \otimes \ket{i}_B$. That's why we call the state
in Eq.~(\ref{eq:bellstate}) the Bell state.

\section{Summary and discussion}

In summary, we have extended the quantum teleportation using quantum states
of distinguishable particles to that using anyonic states of Ising anyons. We have
found a systematic braiding $\mathcal{T}$ that distributes each
pair of the Bell state of two Ising anyons to Alice and Bob, whereby
proposed a general protocol for quantum teleportation based on the Ising anyon
model. We have seen that there is a difference between the protocol for
teleporting the unknown state $\ket{\phi}$ of odd Ising anyons and that of even
Ising anyons. The protocol of the latter is the same as the former except that
we require Bob to perform the braiding $b_0^2$, which changes the parity $c=1$
to $c=0$. This is due to the fact that the structures of Hilbert spaces of odd number
and even number of Ising anyons are different. The total charge of $2N+1$ Ising anyons
is always $\sigma$ while that of $2N$ could be $0$ or $1$, although the dimension
of these two Hilbert spaces are the same. The superselection rules divide
the Hilbert space of $2N$ Ising anyons into two subspaces, which cannot be
transformed into each other by braiding these $2N$ Ising anyons. Thus, the
parity $c$ of $4N$ Ising anyons depends on the parities of $2N$ Ising anyons on
two sides while the parity $c$ of $4N+2$ Ising anyons isn't up to the parities
of $2N+1$ Ising anyons on two sides.

This Ising anyonic teleportation is protected by topology. All that is needed in the
protocol is braiding and measurement. It has been
shown~\cite{PhysRevLett.101.010501, PhysRevLett.98.070401} that the projective
measurements of two Ising anyons can be realized using the interferometry
measurements, in which the target's charge can be inferred from the effect on
the interference of probe anyons braiding around the target's charge.

Although quantum entanglement is well known as a resource consumed in quantum
teleportation, there has been no definitive definition of quantum entanglement
of quantum states of anyons.
Our protocol confirms that the Bell state of Ising anyons [see Eq.~(\ref{eq:bellstate})]
is one of the maximally entangled states. To further define the quantum resource
theory,  we should clarify the definitions of free operations
and free states~\cite{RevModPhys.91.025001} in anyonic systems. Along this direction, our protocol may promote
our understandings of anyonic quantum entanglement as a quantum resource.

In the end, MZM as a reasonable candidate for Ising anyon, has been broadly
searched in the experiments. We hope that our theoretical protocol might be
further studied experimentally in future.

\begin{acknowledgements}
  The authors kindly acknowledge support from National Key Research and
  Development Program of China (Grant No. 2021YFA0718302 and No.
  2021YFA1402104), National Natural Science Foundation of China (Grants No.
  11775300 and No. 12075310), and the Strategic Priority Research Program of
  Chinese Academy of Sciences (Grant No. XDB28000000).
\end{acknowledgements}

\appendix

\onecolumngrid

\section{The Ising anyon model}\label{app:one}

Here we will use the Ising anyon model~\cite{KITAEV20062,bonderson2017anyonic},
a kind of modular tensor categories~\cite{categoryphysics, 2001Lectures},
to derive that all $n$-qubit Pauli gates can be obtained by braiding $2n+2$
or $2n+1$ Ising anyons under the anyonic computational basis.

First, we review the Ising anyon model, which contains three types of topological
charges $\mathcal{I} = \left\{ 0, \sigma, 1 \right\}$ satisfying non-trivial fusion
rules:
\begin{align}
    & \sigma \times \sigma = 0 + 1, \\
    & 1 \times \sigma = \sigma, \\
    & 1 \times 1 = 0,
\end{align}
where $\sigma$ denotes Ising anyon, $1$ denotes fermion, and $0$ is vacuum.

Based on the above fusion rules, we can define the Hilbert space called the fusion
space, which is spanned by the different fusion paths. For example, the fusion
space of two $\sigma$'s fusing into the vacuum is given by
$V_{\sigma^2}^0 = {\rm span} \left\{ \bra{\sigma, \sigma; 0} \right\}$.
Similarly the fusion space of $2$ $\sigma$'s fusing into $1$ is given by
$V_{\sigma^2}^1 = {\rm span} \left\{ \bra{\sigma, \sigma; 1} \right\}$. In particular,
each of these two spaces has only one basis vector since each of them has only one
fusion path. It is useful to employ a diagrammatic representation for anyon
models, where each anyon is associated with an oriented (we will omit the
orientation here) line that can be understood as the anyon's world line.
In the diagrammatic representation, the two basis vectors above can be
represented as
\begin{align}
  \bra{\sigma, \sigma; 0} & = \left( \frac{1}{d_\sigma} \right)^{\frac{1}{2}}
    \begin{tikzpicture}[baseline]
      \draw[dotted] (0,0)--(0,0.5) node[pos=0.5, right]{$0$};
      \draw (0,0)--(0.5,-0.5) node[pos=1, below]{$\sigma$}; 
      \draw (0,0)--(-0.5,-0.5) node[pos=1, below]{$\sigma$};
    \end{tikzpicture}, \nonumber\\
  \bra{\sigma, \sigma; 1} & = \left( \frac{1}{d_\sigma} \right)^{\frac{1}{2}}
    \begin{tikzpicture}[baseline]
      \draw[dotted] (0,0)--(0,0.5) node[pos=0.5, right]{$1$};
      \draw (0,0)--(0.5,-0.5) node[pos=1, below]{$\sigma$}; 
      \draw (0,0)--(-0.5,-0.5) node[pos=1, below]{$\sigma$};
    \end{tikzpicture},
\end{align}
where $d_\sigma = \sqrt{2}$ is the quantum dimension of anyon $\sigma$,
$\left( 1/d_\sigma \right)^{\frac{1}{2}}$ is the normalized coefficient,
the solid line denotes $\sigma$, and the dotted line
denotes the vacuum $0$ and the fermion $1$.

For a system with more $\sigma$'s, the Hilbert space is constructed by taking the tensor
product of its composite parts. For example, the fusion space $V^\sigma_{\sigma^3}$
of three $\sigma$'s with total charge $\sigma$ can be constructed as
\begin{equation}
  V^\sigma_{\sigma^3} \cong \bigoplus_b V^b_{\sigma^2} \otimes V^\sigma_{b, \sigma},
\end{equation}
where $b \in \left\{ 0, 1 \right\}$. It should be noted that fusion order is not
unique. In the example above you can choose to start with fusion of the two $\sigma$'s
on the left or the two $\sigma$'s on the right. These two different methods of fusion
are related by $F$ matrix:
\begin{align}
    \label{eq:fmove}
    \begin{tikzpicture}[baseline=(current bounding box.center)]
      \draw (-1,-1) -- (-0.5,-0.5) node[pos=0, below]{$\sigma$} --(0,-1)
          node[pos=1, below]{$\sigma$};
      \draw (1,-1) -- (0,0) node[pos=0, below]{$\sigma$} --(0,0.5)
          node[pos=0.5, right]{$\sigma$};
      \draw[dotted] (-0.5,-0.5) -- (0,0) node[pos=0.3, above]{$b$};
    \end{tikzpicture} = \sum_d \left( F_\sigma^{\sigma \sigma \sigma} \right)^b_d
    \begin{tikzpicture}[baseline=(current bounding box.center)]
        \draw (-1,-1)--(0,0) node[pos=0, below]{$\sigma$};
        \draw[dotted] (0,0) -- (0.5,-0.5) node[pos=0.8, above]{$d$};
        \draw (0,-1)--(0.5,-0.5) node[pos=0, below]{$\sigma$} -- (1,-1)
            node[pos=1, below]{$\sigma$};
        \draw (0,0)--(0,0.5) node[pos=0.5, right]{$\sigma$};
    \end{tikzpicture},
\end{align}
where $b,d \in \{ 0, 1 \}$ and 
\begin{align}
  F_\sigma^{\sigma \sigma \sigma} = \frac{1}{\sqrt{2}}
    \begin{pmatrix}
      1 & 1 \\
      1 & -1 \\
    \end{pmatrix}.
\end{align}

A linear anyonic operator can be defined using the basis vectors in fusion space and
splitting space as we do in quantum mechanics. For example, the identity operator
for two Ising anyons is
\begin{align}
  \mathbbm{1}_{\sigma \sigma} = & \ket{\sigma, \sigma; 0} \bra{\sigma, \sigma; 0}
    + \ket{\sigma, \sigma; 1} \bra{\sigma, \sigma; 1} \nonumber\\
  = & \frac{1}{d_\sigma} \begin{tikzpicture}[baseline]
    \draw (-0.5,0.75)--(0,0.25)--(0.5,0.75);
    \draw (-0.5,-0.75)--(0,-0.25) --(0.5,-0.75);
  \end{tikzpicture} + \frac{1}{d_\sigma} \begin{tikzpicture}[baseline]
    \draw (-0.5,0.75)--(0,0.25) --(0.5,0.75);
    \draw (-0.5,-0.75)--(0,-0.25)--(0.5,-0.75);
    \draw[dotted] (0,0.25)--(0,-0.25) node[pos=0.5, left]{$1$};
  \end{tikzpicture}.
\end{align}

The quantum trace, which joins the outgoing anyon lines of the anyonic operator
back onto the corresponding incoming lines, e.g.,
\begin{equation}
  \tilde{\rm Tr} \left[ \frac{1}{d_\sigma} \begin{tikzpicture}[baseline]
    \draw (-0.5,0.75)--(0,0.25) --(0.5,0.75);
    \draw (-0.5,-0.75)--(0,-0.25)--(0.5,-0.75);
  \end{tikzpicture} \right] = \frac{1}{d_\sigma}
  \begin{tikzpicture}[baseline]
    \draw (-0.5,0.75)--(0,0.25) --(0.5,0.75);
    \draw (-0.5,-0.75)--(0,-0.25) --(0.5,-0.75);
    \draw (0.5,0.75)--(1,0.25)--(1,-0.25)--(0.5,-0.75);
    \draw (-0.5,0.75)--(0.3,1.55)--(1.2,0.65)--(1.2,-0.65)--(0.3,-1.55)--(-0.5,-0.75);
  \end{tikzpicture} = 1.
\end{equation}
By using the quantum trace, we can define an operator $\tilde{\rho}$, called an anyonic
density operator satisfying the normalization condition
$\tilde{\rm Tr} \left[ \tilde{\rho} \right] = 1$ and the positive semi-definite
condition; that is, for any anyonic state $\ket{\phi}$, we have
$\tilde{\rm Tr} \left[ \bra{\phi} \tilde{\rho} \ket{\phi} \right] \ge 0$.

In addition to fusion rules, the Ising anyon model also needs to meet the rules of
braiding. Specifically, exchanging neighboring $\sigma$'s gives to the anyonic state
a unitary evolution named the $R$ matrix:
\begin{align}
    \begin{tikzpicture}[baseline]
      \draw[dotted] (0,-0.5)--(0,0) node[pos=0.5,right]{$b$};
      \draw (0,0) to [out=160,in=225] (-0.2,0.5)--(0.3,1) node[pos=1,above]{$\sigma$};
      \draw (0,0) to [out=20,in=-45] (0.2,0.5)--(0.1,0.6);
      \draw (-0.1,0.8)--(-0.3,1) node[pos=1,above]{$\sigma$};
    \end{tikzpicture} & = \sum_{d} \left( R_{\sigma \sigma} \right)_{bd}
      \begin{tikzpicture}[baseline]
        \draw[dotted] (0,-0.5)--(0,0) node[pos=0.5,right]{$d$};
        \draw (0,0)--(-0.5,0.5) node[pos=1,above]{$\sigma$};
        \draw (0,0)--(0.5,0.5) node[pos=1,above]{$\sigma$};
      \end{tikzpicture},
\end{align}
where
\begin{equation}
    R_{\sigma \sigma} =
        \begin{pmatrix}
            1 & 0 \\
            0 & i \\
        \end{pmatrix}.
\end{equation}

\section{Pauli gates}\label{app:two}

Suppose there are $2n+1$ Ising anyons in a row. In the fusion order from left to right,
the standard basis $\ket{\overline{p}}$ of this anyonic system can be denoted as
\begin{align}
    \ket{\overline{p}} \equiv \ket{\overline{a}_1 \overline{a}_2 \cdots
        \overline{a}_n; \sigma}_{2n+1}
        = \left( \frac{1}{d_\sigma} \right)^{n/2}
        \begin{tikzpicture}[baseline, scale = 0.8]
            \draw (-3,1.5)--(-2.5,1) node[pos=0,above]{$\sigma_1$}--(-2,1.5)
                node[pos=1,above]{$\sigma_2$};
            \draw (-1,1.5)--(-2,0.5) node[pos=0,above]{$\sigma_3$}--(-1.5,0)--(0,1.5)
                node[pos=1,above]{$\sigma_4$};
            \draw (2,1.5)--(-0.5,-1) node[pos=0,above]{$\sigma_{2n}$};
            \draw (3,1.5)--(0,-1.5) node[pos=0,above]{$\sigma_{2n+1}$}--(0,-2);
            \draw[dotted] (-2.5,1)--(-2,0.5) node[pos=0.6,left]{$\overline{a}_1$};
            \draw[dotted] (-1.5,0)--(-1,-0.5)node[pos=0.6,left]{$\overline{a}_2$};
            \draw (-0.75,-0.75) node[]{$\cdots$};
            \draw (1,1.5) node[]{$\cdots$};
            \draw[dotted] (-0.5,-1)--(0,-1.5) node[pos=0.6,left]{$\overline{a}_n$};
        \end{tikzpicture},
\end{align}
where the solid line denotes the Ising anyon $\sigma$, and the dotted line denotes
two possible fusion results $\overline{a}_i = 0,1$, $i=1,\cdots,n$, encoded as a binary code.
Thus, we see this Hilbert space has dimension $2^{n}$. However, these binary variables
are not independent of each other. To make this basis $\ket{\overline{p}}$ more likes
a $n$-qubit state, we adopt another fusion order, where $\sigma_{2j-1}$ and $\sigma_{2j}$
are fused first. Thus, we give another basis $\ket{p}$ called anyonic computational basis
of the same Hilbert space:
\begin{align}
    \label{eq:barbasis}
    \ket{p} & \equiv \ket{a_1 a_1 \cdots a_n; \sigma}_{2n+1}
        \nonumber\\
    & = \left( \frac{1}{d_\sigma} \right)^{n/2}
        \begin{tikzpicture}[baseline, scale = 0.8]
            \draw (-3.5,1.5)--(-3,1) node[pos=0,above]{$\sigma_1$}--(-2.5,1.5)
                node[pos=1,above]{$\sigma_2$};
            \draw (-1.5,1.5)--(-1,1) node[pos=0,above]{$\sigma_3$}--(-0.5,1.5)
                node[pos=1,above]{$\sigma_4$};
            \draw (1.5,1.5)--(2,1) node[pos=0,above]{$\sigma_{2n-1}$}--(2.5,1.5)
                node[pos=1,above]{$\sigma_{2n}$};
            \draw (0.5,1.5) node[]{$\cdots$};
            \draw[dotted] (-3,1)--(-3,0.3) node[pos=0.5,right]{$a_1$};
            \draw[dotted] (-1,1)--(-1,0.3) node[pos=0.5,right]{$a_2$};
            \draw[dotted] (2,1)--(2,0.3) node[pos=0.5,right]{$a_n$};
            \draw (3.5,1.5)--(3.5,0.3) node[pos=0,above]{$\sigma_{2n+1}$};
            \draw (-3.7,-1) rectangle (3.7,0.3);
            \draw (0,-1)--(0,-1.8);
            \draw (0,-0.4) node[]{Fusion path of $a_1 \times \cdots \times
                a_n \times \sigma = \sigma$};
        \end{tikzpicture},
\end{align}
where $a_i = 0,1$, $i=1,\cdots,n$. The fusion path in the rectangle box above is not
showed, since it can be uniquely determined due to abelian fusion rules of $0$ and $1$.

These two bases span the same Hilbert space are related by trivial $F$ moves in the
Ising anyon model, for example, when $n=2$, we have
\begin{align}
    \ket{\overline{0} \overline{0}; \sigma}_5 = \ket{00; \sigma}_5, ~~~~
    \ket{\overline{0} \overline{1}; \sigma}_5 = \ket{01; \sigma}_5,~~~~
    \ket{\overline{1}\overline{0}; \sigma}_5 = \ket{11; \sigma}_5, ~~~~
    \ket{\overline{1}\overline{1}; \sigma}_5 = \ket{10; \sigma}_5.
\end{align}
We can view these $F$ moves as a CNOT gate $U_{\rm CN}^{1,2}$ with $\overline{a}_1$
as the control qubit and $\overline{a}_2$ as the target qubit. Thus, we can get the
basis $\ket{p}$ from the basis $\ket{\overline{p}}$ using a series of CNOT gates:
\begin{align}
    \ket{a_1, a_1, \cdots, a_n; \sigma}_{2n+1} & = U_{\rm CN}^{1,2}
        U_{\rm CN}^{2,3} \cdots U_{\rm CN}^{n-1,n} \ket{\overline{a}_1, \overline{a}_2,
        \cdots, \overline{a}_n; \sigma}_{2n+1} \nonumber\\
    & \equiv U_{\rm CNOT} \ket{\overline{a}_1, \overline{a}_2, \cdots,
        \overline{a}_n; \sigma}_{2n+1}
\end{align}

Now we consider the states $\ket{p}$ as the basis for our representation space.
We will use the $F$ move and the $R$ matrix to give a representation to braid group.
Explicitly, we have the generators $b_{j}^{(2n+1)}$ of braid group $B_{2n+1}$,
which denotes the exchange of the $j$-th and $(j+1)$-th Ising anyons in the defined
positive direction,
\begin{align}
    \label{eq:2n1braid}
    b^{(2n+1)}_{2j-1} & = \underbrace{ I_2 \otimes \cdots \otimes I_2 }_{j-1} \otimes
        \begin{pmatrix}
            1 & 0 \\
            0 & i \\
        \end{pmatrix} \otimes \underbrace{ I_2 \otimes \cdots \otimes I_2 }_{n-j},
            ~~~~~~~~~~~~~~~~~~~~~~~~~ 1\le j \le n,
        \nonumber \\
    b^{(2n+1)}_{2j} & = U_{\rm CN}^{j,j+1} ( F^{\sigma\sigma\sigma}_{\sigma} \otimes I_2 )
        (R_{\sigma \sigma} \otimes I_2) ( F^{\sigma\sigma\sigma}_{\sigma} \otimes I_2 )
        U_{\rm CN}^{j,j+1} \nonumber\\
    & = \underbrace{ I_2 \otimes \cdots \otimes I_2 }_{j-1} \otimes
        \frac{{\rm e}^{i\pi/4}}{\sqrt{2}}
        \begin{pmatrix}
            1 & 0 & 0 & -i \\
            0 & 1 & -i & 0 \\
            0 & -i & 1 & 0 \\
            -i & 0 & 0 & 1 \\
        \end{pmatrix} \otimes \underbrace{ I_2 \otimes \cdots \otimes I_2 }_{n-j-1},~~~~
        1 \le j \le n-1, \nonumber \\
    b^{(2n+1)}_{2n} & = \underbrace{ I_2 \otimes \cdots \otimes I_2 }_{n-1}
        \otimes \frac{{\rm e}^{i\pi/4}}{\sqrt{2}}
        \begin{pmatrix}
            1 & -i \\
            -i & 1 \\
        \end{pmatrix},
\end{align}
where $b^{(2n+1)}_{2j}$, $1 \le j \le n-1$, can be proved by noticing that
\begin{align}
    \begin{tikzpicture}[baseline, scale = 0.8]
        \draw (-1.5,1.5) node[above]{$\sigma_{2j-1}$}--(-1,1)--(-0.5,1.5)
            node[above]{$\sigma_{2j}$};
        \draw (1.5,1.5) node[above]{$\sigma_{2j+2}$}--(1,1)--(0.5,1.5)
            node[above]{$\sigma_{2j+1}$};
        \draw[dotted] (-1,1)--(0,0) node[pos=0.3,below]{$a_j$}--(1,1);
        \draw (1,0.3) node[]{$a_{j+1}$};
        \draw[dotted] (0,0)--(0,-1) node[pos=0.5,right]{$a'$};
    \end{tikzpicture}
        \overset{ U_{CN}^{j,j+1} }{\Rightarrow}
        \begin{tikzpicture}[baseline, scale = 0.8]
            \draw (-1.5,1.5) node[above]{$\sigma_{2j-1}$}--(-1,1)--(-0.5,1.5)
                node[above]{$\sigma_{2j}$};
            \draw (0.5,1.5) node[above]{$\sigma_{2j+1}$}--(-0.5,0.5)--(0,0)--(1.5,1.5)
                node[above]{$\sigma_{2j+2}$};
            \draw[dotted] (-1,1)--(-0.5,0.5) node[pos=0.3,below]{$\overline{a}_j$};
            \draw[dotted] (0,0)--(0,-1) node[pos=0.5,right]{$\overline{a}'$};
        \end{tikzpicture}
        \overset{F^{\sigma\sigma\sigma}_{\sigma}}{\Rightarrow}
    \begin{tikzpicture}[baseline, scale = 0.8]
        \draw (-1.5,1.5) node[above]{$\sigma_{2j-1}$}--(0,0);
        \draw (0.5,1.5) node[above]{$\sigma_{2j+1}$}--(0,1)--(-0.5,1.5)
            node[above]{$\sigma_{2j}$};
        \draw (1.5,1.5) node[above]{$\sigma_{2j+2}$}--(0,0);
        \draw[dotted] (0,0)--(0,-1) node[pos=0.5,right]{$\overline{a}'$};
        \draw[dotted] (0,1)--(-0.5,0.5);
        \draw (0,0.5) node[]{$\overline{a}_{j}'$};
    \end{tikzpicture}.
\end{align}

Now we have the generators $b_{j}^{(2n+1)}$ of braid group $B_{2n+1}$.
It's straightforward to extend to the case of $B_{2n+2}$:
\begin{align}
    \label{eq:2n2braid}
    b^{(2n+2)}_{2j-1} & = \underbrace{ I_2 \otimes \cdots \otimes I_2 }_{j-1} \otimes
        \begin{pmatrix}
            1 & 0 \\
            0 & i \\
        \end{pmatrix} \otimes \underbrace{ I_2 \otimes \cdots \otimes I_2 }_{n-j+1},
        ~~~~~~~~~~~~~~~~~~~~~~~~~ 1\le j \le n+1, \nonumber \\
    b^{(2n+2)}_{2j} & = \underbrace{ I_2 \otimes \cdots \otimes I_2 }_{j-1} \otimes
        \frac{{\rm e}^{i\pi/4}}{\sqrt{2}}
        \begin{pmatrix}
            1 & 0 & 0 & -i \\
            0 & 1 & -i & 0 \\
            0 & -i & 1 & 0 \\
            -i & 0 & 0 & 1 \\
        \end{pmatrix} \otimes \underbrace{ I_2 \otimes \cdots \otimes I_2 }_{n-j},~~~~
        1 \le j \le n.
\end{align}
These matrices with dimension $2^{n+1} \times 2^{n+1}$ are reducible due to the
superselection rule. In other words, the parity $\mathcal{P}$ of
the state $\ket{p} = \ket{a_1 \cdots a_{n+1}}$,
which is defined as $\mathcal{P} = \sum_i a_i$ mod $2$, gives two representations
for braid group $B_{2n+2}$. Indeed, we see that $\tau_3 ^{\otimes (n+1)}$ commutes with
all generators of $B_{2n+2}$, where $\tau_i$, $i=0,1,2,3$, are Pauli matrices.
Since the eigenvalues of $\tau_3 ^{\otimes (n+1)}$ have two different values $\pm$,
this reducible representation can be reduced into two irreducible representations
by projectors
\begin{equation}
    \label{eq:projector}
    P_\pm^{(2n+2)} = \frac{I \pm \tau_3 ^{\otimes (n+1)}}{2},
\end{equation}
where the superscript $\pm$ denotes two irreducible representations. The generators
of $B_{2n+2}$ derived here are the same as the results in the Ref.~\cite{Georgiev_2009}.

Finally, all $n$-qubit Pauli gates can be obtained by braiding $2n+1$ Ising anyons
in the computational basis with $2n+1$ Ising anyons:
\begin{align}
    & \tau^{(j)}_3 = \left( b^{(2n+1)}_{2j-1} \right)^2, \nonumber\\
    & \tau_1^{(j)} = \left( b^{(2n+1)}_{2j} \right)^2 \left( b^{(2n+1)}_{2j+2} \right)^2
        \cdots \left( b^{(2n+1)}_{2n} \right)^2, \nonumber \\
    & i = b^{(2n+1)}_{2n-1} \left( b^{(2n+1)}_{2n} \right)^2 b^{(2n+1)}_{2n-1}
        \left( b^{(2n+1)}_{2n} \right)^2, \nonumber
\end{align}
or by braiding $2n+1$ Ising anyons in the computational basis with $2n+2$ Ising anyons:
\begin{align}
  \label{eq:PGeven}
    & \tau_3^{(j)} = \left( b^{(2n+2)}_{2j-1} \right)^2, \nonumber \\
    & \tau^{(j)}_1 \otimes \tau_1^{(n+1)} = \left( b^{(2n+2)}_{2j} \right)^2 \left(
        b^{(2n+2)}_{2j+2} \right)^2 \cdots \left( b^{(2n+2)}_{2n} \right)^2, \\
    & i = b^{(2n+2)}_{2n-1} \left( b^{(2n+2)}_{2n} \right)^2 b^{(2n+2)}_{2n-1}
    \left( b^{(2n+2)}_{2n} \right)^2, \nonumber
\end{align}
where $1 \le j \le n$, and $\tau_i^{(j)}$ is Pauli matrix $\tau_i$ acting on
the $j$-th qubit $a_j$. It is worth noting that, in simulating $n$-qubit Pauli gates
in the computational basis with $2n+1$ and $2n+2$ Ising anyons, we use the same
braiding operations on $2n+1$ Ising anyons. However,
for braiding $2n+2$ Ising anyons, if we want to apply $\tau_1^{(j)}$
or $\tau_2^{(j)}$ on qubit $a_j$, it will not only change the $a_j$ but will change
the assistant qubit $a_{n+1}$ to maintain the parity.

\section{Quantum tracing Alice's side}\label{app:three}

In this section, we will give details of how to obtain the state $\ket{\phi}$
of $M$ (odd) Ising anyons from the state in Eq.~(\ref{eq:16}) by quantum tracing
charge $c \in \left\{ 0,1 \right\}$, which is the parity of Alice's measurement
outcomes, i.e., the total charge of the state obtained by Alice's measurement:
\begin{align}
  \label{eq:c1}
  \tilde{\rm Tr}_c \left[
  \begin{tikzpicture}[baseline=(current bounding box.center), scale = 0.6]
    \draw (-0.2,0) rectangle (1.2,1);
    \draw (0.5,0.5) node[]{$\ket{\psi}$};
    \draw (-0.3,1.2) -- (-1,0.5) -- (0,-0.5) -- (0,-1.5)
      -- (-1,-2.5) -- (-0.3,-3.2);
    \draw[dotted] (0.5,0) -- (0,-0.5) node[pos=0.2, below]{$i$};
    \draw[dotted] (-1.7,1.2) -- (-1,0.5) node[pos=0, above]{$c$};
    \draw (0,1) -- (0,1.2);
    \draw (1,1) -- (1,1.2);
    \draw[dotted] (0.3,1.2) -- (0.7,1.2);
    \draw[dotted] (-1,-2.5) -- (-1.7,-3.2) node[below]{$c$};
    \draw[dotted] (0,-1.5) -- (0.5,-2) node[above]{$i$};
    \draw (-0.2,-2) rectangle (1.2,-3);
    \draw (0.5,-2.5) node[]{$\ket{\psi}$};
    \draw (0,-3) -- (0,-3.2);
    \draw (1,-3) -- (1,-3.2);
    \draw[dotted] (0.3,-3.2) -- (0.7,-3.2);
  \end{tikzpicture} \right] = \ket{\phi} \bra{\phi},
\end{align}
where $\ket{\psi}$ denotes the state of $M-1$ (even) Ising anyons,
and $\tilde{\rm Tr}_c$ denotes partial quantum trace over charge $c$.
We use anyonic density matrices for anyonic states in order to apply
quantum trace directly. From the left-hand side of Eq.~(\ref{eq:c1}),
we have:
\begin{align}
  \begin{tikzpicture}[baseline=(current bounding box.center), scale = 0.6]
    \draw (-0.2,0) rectangle (1.2,1);
    \draw (0.5,0.5) node[]{$\ket{\psi}$};
    \draw (-0.3,1.2) -- (-1,0.5) -- (0,-0.5) -- (0,-1.5)
      -- (-1,-2.5) -- (-0.3,-3.2);
    \draw[dotted] (0.5,0) -- (0,-0.5) node[pos=0.2, below]{$i$};
    \draw[dotted] (-1.7,1.2) -- (-1,0.5) node[pos=0, above]{$c$};
    \draw (0,1) -- (0,1.2);
    \draw (1,1) -- (1,1.2);
    \draw[dotted] (0.3,1.2) -- (0.7,1.2);
    \draw[dotted] (-1,-2.5) -- (-1.7,-3.2) node[below]{$c$}
      --(-2.4,-2.5) -- (-2.4,0.5) -- (-1.7,1.2);
    \draw[dotted] (0,-1.5) -- (0.5,-2) node[above]{$i$};
    \draw (-0.2,-2) rectangle (1.2,-3);
    \draw (0.5,-2.5) node[]{$\ket{\psi}$};
    \draw (0,-3) -- (0,-3.2);
    \draw (1,-3) -- (1,-3.2);
    \draw[dotted] (0.3,-3.2) -- (0.7,-3.2);
  \end{tikzpicture} \quad = \quad
  \begin{tikzpicture}[baseline=(current bounding box.center), scale = 0.6]
    \draw (0.3,0.5) rectangle (1.7,1.5);
    \draw (1,1) node[]{$\ket{\psi}$};
    \draw (-0.5,1.7) -- (-0.5,1) -- (0.5,0) -- (0,-0.5) -- (0,-1.5)
      -- (0.5,-2) -- (-0.5,-3) -- (-0.5,-3.7);
    \draw[dotted] (1,0.5) -- (0.5,0) node[pos=0.2, below]{$i$};
    \draw[dotted] (-1.7,1.2) -- (0,-0.5) node[pos=0, above]{$c$};
    \draw (0.5,1.5) -- (0.5,1.7);
    \draw (1.5,1.5) -- (1.5,1.7);
    \draw[dotted] (0.8,1.7) -- (1.2,1.7);
    \draw[dotted] (0,-1.5) -- (-1,-2.5) -- (-1.7,-3.2) node[below]{$c$}
      --(-2.4,-2.5) -- (-2.4,0.5) -- (-1.7,1.2);
    \draw[dotted] (0.5,-2) -- (1,-2.5) node[above]{$i$};
    \draw (0.3,-2.5) rectangle (1.7,-3.5);
    \draw (1,-3) node[]{$\ket{\psi}$};
    \draw (0.5,-3.5) -- (0.5,-3.7);
    \draw (1.5,-3.5) -- (1.5,-3.7);
    \draw[dotted] (0.8,-3.7) -- (1.2,-3.7);
  \end{tikzpicture} \quad = \quad
  \begin{tikzpicture}[baseline=(current bounding box.center), scale = 0.6]
    \draw (0.3,0.5) rectangle (1.7,1.5);
    \draw (1,1) node[]{$\ket{\psi}$};
    \draw (-0.5,1.7) -- (-0.5,1) -- (0.5,0) -- (0.5,-2) -- (-0.5,-3) -- (-0.5,-3.7);
    \draw[dotted] (1,0.5) -- (0.5,0) node[pos=0.2, below]{$i$};
    \draw[dotted] (-1.7,1.2) -- (-1,0.5) node[pos=0, above]{$c$}
      -- (-1,-2.5);
    \draw (0.5,1.5) -- (0.5,1.7);
    \draw (1.5,1.5) -- (1.5,1.7);
    \draw[dotted] (0.8,1.7) -- (1.2,1.7);
    \draw[dotted] (-1,-2.5) -- (-1.7,-3.2) node[below]{$c$}
      --(-2.4,-2.5) -- (-2.4,0.5) -- (-1.7,1.2);
    \draw[dotted] (0.5,-2) -- (1,-2.5) node[above]{$i$};
    \draw (0.3,-2.5) rectangle (1.7,-3.5);
    \draw (1,-3) node[]{$\ket{\psi}$};
    \draw (0.5,-3.5) -- (0.5,-3.7);
    \draw (1.5,-3.5) -- (1.5,-3.7);
    \draw[dotted] (0.8,-3.7) -- (1.2,-3.7);
  \end{tikzpicture} \quad = \quad \ket{\phi} \bra{\phi}.
\end{align}
In the first step we have used $F$-move and the $F$ matrix $F^{c\sigma i}_\sigma$
is a number. In the second step we have used $F$-move with two lower and
two upper legs and the fact that the tadpole diagram gives zero. In the last step
we have used the fact that an unknotted loop carrying charge $c$ gives to its
quantum dimension $d_c$.

From Eqs.~(\ref{eq:16}) and (\ref{eq:c1}), we see that, when Alice's
measurement outcomes are $\left\{ 10 \cdots 0 \right\}$, the state of $M$ (odd)
Ising anyons on Bob's side is $\ket{\phi}$ just like the case where the measurement
outcomes $\left\{ 0\cdots 0 \right\}$.

However, this does not apply when $M$ is even:
\begin{align}
  \begin{tikzpicture}[baseline=(current bounding box.center), scale = 0.6]
    \draw (-0.2,0) rectangle (1.2,1);
    \draw (0.5,0.5) node[]{$\ket{\psi}$};
    \draw (-0.3,1.2) -- (-1,0.5) -- (0,-0.5);
    \draw[dotted] (0,-0.5) -- (0,-1.5) node[pos=0.5, right]{$i$};
    \draw (0,-1.5) -- (-1,-2.5) -- (-0.3,-3.2);
    \draw (0.5,0) -- (0,-0.5);
    \draw[dotted] (-1.7,1.2) -- (-1,0.5) node[pos=0, above]{$c$};
    \draw (0,1) -- (0,1.2);
    \draw (1,1) -- (1,1.2);
    \draw[dotted] (0.3,1.2) -- (0.7,1.2);
    \draw[dotted] (-1,-2.5) -- (-1.7,-3.2) node[below]{$c$}
      --(-2.4,-2.5) -- (-2.4,0.5) -- (-1.7,1.2);
    \draw (0,-1.5) -- (0.5,-2);
    \draw (-0.2,-2) rectangle (1.2,-3);
    \draw (0.5,-2.5) node[]{$\ket{\psi}$};
    \draw (0,-3) -- (0,-3.2);
    \draw (1,-3) -- (1,-3.2);
    \draw[dotted] (0.3,-3.2) -- (0.7,-3.2);
  \end{tikzpicture} \quad = \quad
  \begin{tikzpicture}[baseline=(current bounding box.center), scale = 0.6]
    \draw (0.3,0.5) rectangle (1.7,1.5);
    \draw (1,1) node[]{$\ket{\psi}$};
    \draw (-0.5,1.7) -- (-0.5,1) -- (0.5,0);
    \draw[dotted] (0.5,0) -- (0,-0.5) node[pos=0.5, right]{$i-c$};
    \draw[dotted] (0,-0.5) -- (0,-1.5) node[pos=0.5, right]{$i$};
    \draw[dotted] (0,-1.5) -- (0.5,-2) node[pos=0.5, right]{$i-c$};
    \draw (1,0.5) -- (0.5,0);
    \draw[dotted] (-1.7,1.2) -- (0,-0.5) node[pos=0, above]{$c$};
    \draw (0.5,1.5) -- (0.5,1.7);
    \draw (1.5,1.5) -- (1.5,1.7);
    \draw[dotted] (0.8,1.7) -- (1.2,1.7);
    \draw[dotted] (0,-1.5) -- (-1,-2.5) -- (-1.7,-3.2) node[below]{$c$}
      --(-2.4,-2.5) -- (-2.4,0.5) -- (-1.7,1.2);
    \draw (-0.5,-3.7) -- (-0.5,-3) -- (0.5,-2) -- (1,-2.5);
    \draw (0.3,-2.5) rectangle (1.7,-3.5);
    \draw (1,-3) node[]{$\ket{\psi}$};
    \draw (0.5,-3.5) -- (0.5,-3.7);
    \draw (1.5,-3.5) -- (1.5,-3.7);
    \draw[dotted] (0.8,-3.7) -- (1.2,-3.7);
  \end{tikzpicture} \quad = \quad
  \begin{tikzpicture}[baseline=(current bounding box.center), scale = 0.6]
    \draw (0.3,0.5) rectangle (1.7,1.5);
    \draw (1,1) node[]{$\ket{\psi}$};
    \draw (-0.5,1.7) -- (-0.5,1) -- (0.5,0);
    \draw (0.5,-2) -- (-0.5,-3) -- (-0.5,-3.7);
    \draw[dotted] (0.5,0) -- (0.5,-2) node[pos=0.5, right]{$i-c$};
    \draw (1,0.5) -- (0.5,0);
    \draw[dotted] (-1.7,1.2) -- (-1,0.5) node[pos=0, above]{$c$}
      -- (-1,-2.5);
    \draw (0.5,1.5) -- (0.5,1.7);
    \draw (1.5,1.5) -- (1.5,1.7);
    \draw[dotted] (0.8,1.7) -- (1.2,1.7);
    \draw[dotted] (-1,-2.5) -- (-1.7,-3.2) node[below]{$c$}
      --(-2.4,-2.5) -- (-2.4,0.5) -- (-1.7,1.2);
    \draw (0.5,-2) -- (1,-2.5);
    \draw (0.3,-2.5) rectangle (1.7,-3.5);
    \draw (1,-3) node[]{$\ket{\psi}$};
    \draw (0.5,-3.5) -- (0.5,-3.7);
    \draw (1.5,-3.5) -- (1.5,-3.7);
    \draw[dotted] (0.8,-3.7) -- (1.2,-3.7);
  \end{tikzpicture} ~,
\end{align}
where $\ket{\psi}$ denotes the state of $M-1$ (odd) Ising anyons. Similarly,
by quantum tracing the charge $c$, we obtain the state of $M$ Ising anyons
with parity $i-c$.
When $c=0$, this state becomes the state $\ket{\phi}$ Alice wants to teleport.
When $c=1$, this state has a different total charge than $\ket{\phi}$, and
we need auxiliary anyon to change the parity.

\twocolumngrid

\bibliographystyle{apsrev4-2} \bibliography{IATbib.bib}

\end{document}